\documentclass{aa}
\usepackage[utf8]{inputenc}

\title{Time-dependent photoionization spectroscopy of the Seyfert galaxy NGC~3783}
\author{Liyi Gu \inst{1,2,3}, Jelle Kaastra \inst{1,3}, Daniele Rogantini \inst{4}, Missagh Mehdipour \inst{5}, Anna Jur$\rm \acute{a}\breve{n}$ov$\rm \acute{a}$ \inst{1,6}, Elisa Costantini \inst{1,6}, and Chen Li \inst{3,1} }
\date{March 2023}

\institute{
SRON Netherlands Institute for Space Research, Niels Bohrweg 4, 2333 CA Leiden, the Netherlands
\and
RIKEN High Energy Astrophysics Laboratory, 2-1 Hirosawa, Wako, Saitama 351-0198, Japan
\and 
Leiden Observatory, Leiden University, PO Box 9513, 2300 RA Leiden, The Netherlands
\and 
MIT Kavli Institute for Astrophysics and Space Research, Massachusetts Institute of Technology, Cambridge, MA 02139, USA
\and 
Space Telescope Science Institute, 3700 San Martin Drive, Baltimore, MD 21218, USA
\and 
Anton Pannekoek Institute, University of Amsterdam, Postbus 94249, NL-1090 GE Amsterdam, the Netherlands
}

\abstract{
We present an investigation into the spectroscopic properties of non-equilibrium photoionization processes operating in a time-evolving mode. Through a quantitative comparison between equilibrium and time-evolving models, we find that the time-evolving model exhibits a broader distribution of charge states compared to the equilibrium model, accompanied by a slight shift in the peak ionization state depending on the source variability and gas density. The time-evolving code, {\tt tpho} in SPEX, has been successfully employed to analyze the spectral properties of warm absorbers in the Seyfert galaxy NGC~3783. The incorporation of variability in the {\tt tpho} model improves the fits of the time-integrated spectra, providing more accurate descriptions to the average charge states of several elements, in particular for Fe which is peaked around \ion{Fe}{XIX}. The inferred densities and distances of the relevant X-ray absorber components are estimated to be approximately a few $10^{11}$ m$^{-3}$ and $\leq 1$~pc, respectively. Furthermore, the updated fit suggests a potential scenario in which the observed absorbers are being expelled from the central AGN at the escape velocities. This implies that these absorbers might not play a significant role in the AGN feedback mechanism. 
}

\titlerunning{TPHO spectroscopy of NGC~3783}
\authorrunning{L. Gu}

\begin{document}

\maketitle

\section{Introduction}

Warm absorbers are among one of the manifestations of photoionized outflows from active galactic nuclei (AGNs). Within the X-ray regime, they are often recognized as absorption features characterized by a diverse range of ionization states. These features typically exhibit hydrogen column densities ranging from $10^{24}$ to $10^{26}$ m$^{-2}$, and outflow velocities of approximately $\sim 100-1000$ km s$^{-1}$ \citep{reynolds1997, blustin2005, k2012, laha2021}. While warm absorbers have been extensively observed and studied, two outstanding questions remain unanswered in the interpretation of these observations: (1) is the conventional practice of modeling absorbers as discrete ionization components, each in photoionization equilibrium, a reasonable approximation? (2) What are the precise locations and mechanisms responsible for launching these outflows?

For warm absorbers, the photoionization equilibrium with an external ionizing source requires a certain duration for it to develop. This timescale is commonly approximated by the recombination time of the absorbers \citep{krolik1995, nicastro1999}. To maintain steady equilibrium, it is crucial for the recombination time to be considerably shorter than the characteristic variation time of the ionizing source. This condition ensures that the system can promptly and effectively respond to fluctuations in the intensity of the ionizing radiation. However, it is not guaranteed that all warm absorbers satisfy this condition. The recombination time depends on factors including electron density and ionization state, and with the current limited knowledge of absorber density (e.g., \citealt{gabel2005, arav2015}), the recombination time could range from $10^{4}$ to $10^{8}$ seconds. On the other hand, power spectral analysis conducted by \citet{u2002} and \citet{m2003} has shown that a significant portion of X-ray variations occur on timescales of $10^{5}$ to $10^{7}$ seconds. When the two timescales become comparable, warm absorbers are unable to achieve equilibrium, resulting in variations in their ionization states lagging behind the intrinsic variations in the AGNs (e.g., \citealt{silva2016}).

To address the non-equilibrium photoionization condition, several time-evolving photoionization codes have recently been developed, building upon existing equilibrium models (\citealt{hoof2020} with CLOUDY; \citealt{sadaula2023} with XSTAR; \citealt{luminari2022} TEPID; and \citealt{r2022} with SPEX). These codes incorporate ionizing lightcurves and spectral energy distributions (SEDs), and solve self-consistently the time-dependent equations governing the ionization and thermal states as a function of gas density. By employing such time-evolving models, a more accurate approximation of the true state of warm absorbers can be obtained, which solves the question (1). Furthermore, these models have the potential to provide insights into question (2), namely the determination of the distances at which the warm absorbers are located. The distance can be constrained using the relation $R = \sqrt{ L / n_{\rm H} \xi}$, where $L$ represents the ionizing luminosity, $n_{\rm H}$ denotes the hydrogen number density, and $\xi$ is the ionization parameter. The time-evolving calculations are able to incorporate all three parameters, whereas in equilibrium modeling, it becomes challenging to constrain the gas density accurately \citep{k2004,k2012}.

A significant portion of previous works has focused on density diagnostics utilizing response lags \citep{nicastro1999,silva2016,j2022,r2022,l2023} and absorption line features from metastable levels \citep{k2004, gabel2005, edmonds2011, digesu2013, m2017}. However, there is only limited knowledge about the overall characteristic spectroscopic features specific to time-evolving photoionized plasmas \citep{r2022}. Acquiring such knowledge would benefit in two aspects: firstly, it would provide a systematic understanding of the practical differences between equilibrium and time-evolving models; secondly, it would allow us to explore the potential to constrain the density and distance of the observed warm absorbers through direct spectral analysis. In this work, we aim to characterize the time-evolving spectral model by differentiating it from the equilibrium model, and apply it to real warm absorber observations. We utilize the SPEX {\tt tpho} \citep{r2022} and {\tt pion} \citep{m2016} models as the baseline for the time-dependent and equilibrium conditions, respectively. 

The time-dependent photoionization model is applied to the existing data of the Seyfert~1 galaxy NGC~3783 ($z$ = 0.009730, \citealt{t1998}). This AGN is well-known for its prominent warm absorber, characterized by narrow absorption lines originating from various elements including Fe, Ca, Ar, S, Si, Al, Mg, Ne, O, N, and C. The observed absorption lines span a wide range of ionization states, with a total column density of $\sim 4 \times 10^{26}$ m$^{-2}$. Previous studies have reported an average outflow velocity of approximately 600~km s$^{-1}$ for this warm absorber \citep{kaspi2000, kaspi2001, kaspi2002, scott2014}. The X-ray absorption spectra have been modeled using different approaches, incorporating two ionized components \citep{blustin2002, krongold2003, krongold2005}, three components \citep{netzer2003}, five components \citep{ballhausen2023}, nine components \citep{m2017, m2019, k2018}, or a broad continuous absorption measure distribution \citep{goncalves2006, behar2009, goosmann2016, keshet2022}. All existing models assume photoionization equilibrium.

The structure of the paper is as follows. In \S~\ref{sec:1}, we present a theoretical exploration of the spectroscopic characteristics of the $\tt tpho$ model. In \S~\ref{sec:app}, we employ the {\tt tpho} model to analyze the {\it Chandra} grating data of NGC~3783. Finally, the findings are discussed and summarized in \S~\ref{sec:3}. Throughout the paper, the 
errors are given at a 68\% confidence level.

\section{Time-dependent photoionization model}
\label{sec:1}
In order to investigate the general spectral characteristics of a time-varying photoionized source, we perform a series of time-dependent calculations using the SPEX {\tt tpho} routine \citep{r2022} and the NGC~3783 ionized outflow configuration as described in \citet{m2017} and \citet{m2019}. By utilizing the eigenvector method initially described in \citet{k1993}, the {\tt tpho} model calculates self-consistently the out-of-equilibrium time evolution of photoionization, accurately determining the ion concentrations and electron temperature of the ionized source as a function of time. One advantage of the {\tt tpho} model, as highlighted in \citet{r2022} and \citet{l2023}, is its potential as a density diagnostic tool for the target source. The recombination process, influenced by collisional interactions, is directly linked to the electron density of the source.

The ionizing SED and its corresponding lightcurve, as well as the NGC~3783 warm absorber outflows, have been modeled in accordance with the methods outlined in \citet{l2023}. The SED comes originally from the time-averaged spectral analysis of the optical to hard X-ray data in its unobscured phase during 2000-2001. As described in \citet{m2017}, the archival data from Hubble space telescope, {\it Chandra}, {\it XMM-Newton}, and {\it NuSTAR} were fit jointly to determine the SED. The lightcurve is derived by a simulation based on the power spectrum obtained from a large set of {\it XMM-Newton} and {\it Rossi} X-ray Timing Explorer (RXTE) data \citep{m2005}. It has been demonstrated in \citet{l2023} that the simulated lightcurve spanning approximately 1$\times 10^{7}$ seconds is adequate for capturing the primary variability that occurs within the range of 10$^5$ to 10$^6$ seconds. The fractional variability of the model lightcurve, defined as the ratio of the standard deviation to the average flux, is approximately 40\%. The warm absorbers in the target are taken from the results of \citet{m2017} and \cite{m2019} using data from the {\it Chandra} and {\it XMM-Newton} gratings, which revealed nine distinct photoionization components with the ionization parameter log $\xi$ spanning a range from -0.65 to 3 (see Table~3 in \citealt{m2019} and Table~\ref{tab_pion} later in this paper). 

\begin{figure}[!htbp]
\centering
\resizebox{1.0\hsize}{!}{\includegraphics[angle=0]{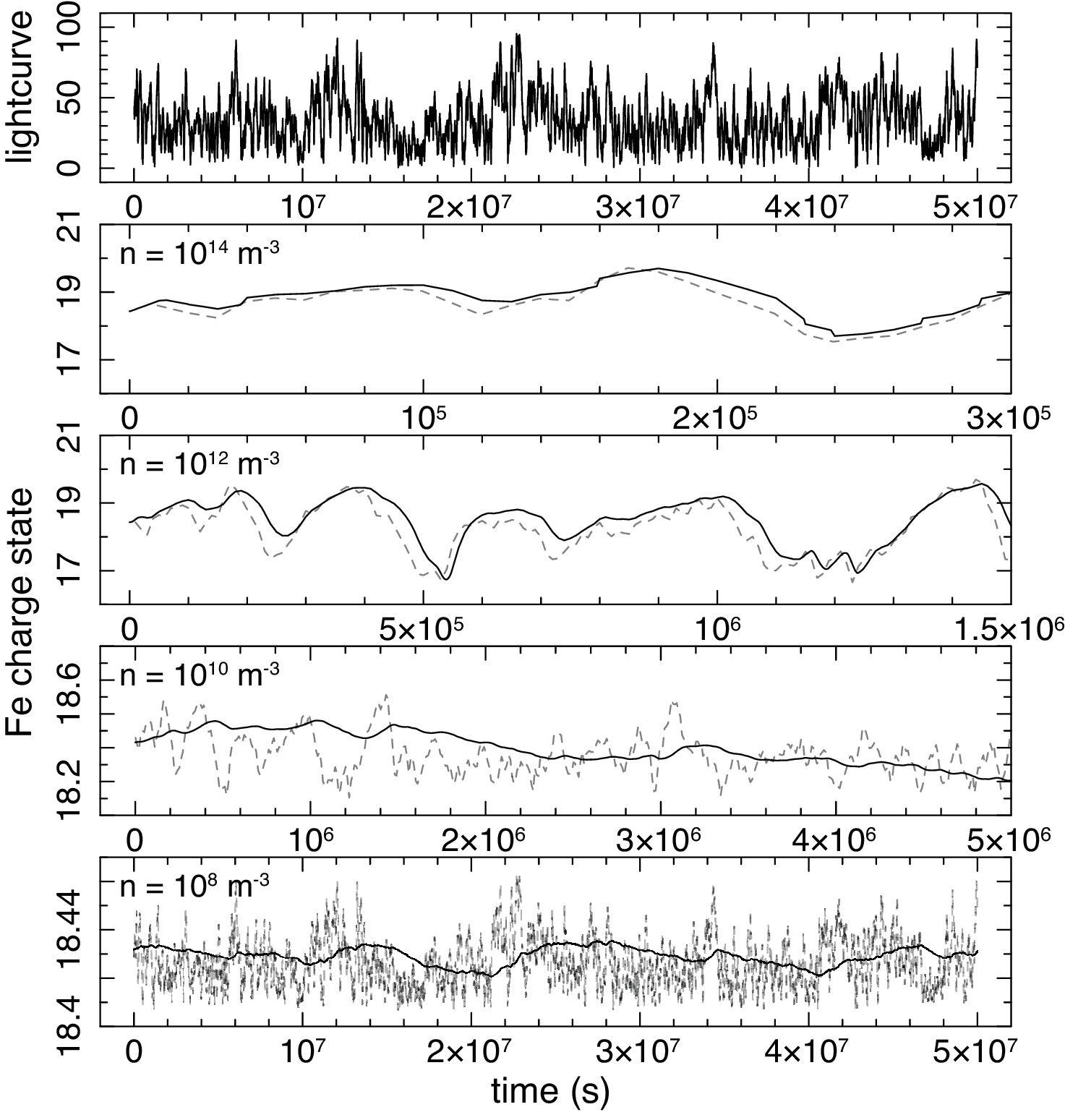}}
\caption{Simulated lightcurve (upper panel) and the corresponding evolution of Fe average charge of component~3 in the panels below. The average charge variations are shown by the solid lines for gas densities of $10^{14}$, $10^{12}$, $10^{10}$, $10^{8}$ m$^{-3}$, while the dashed curves illustrate the corresponding source lightcurves in the same period. As the gas density decreases, the response time to source variations becomes longer, resulting in flatter average charge variations. To highlight the response time differences across various densities, each panel presents a different total time period.  }
\label{fig:lclong}
\end{figure}

\begin{figure}[!htbp]
\centering
\resizebox{1.0\hsize}{!}{\includegraphics[angle=0]{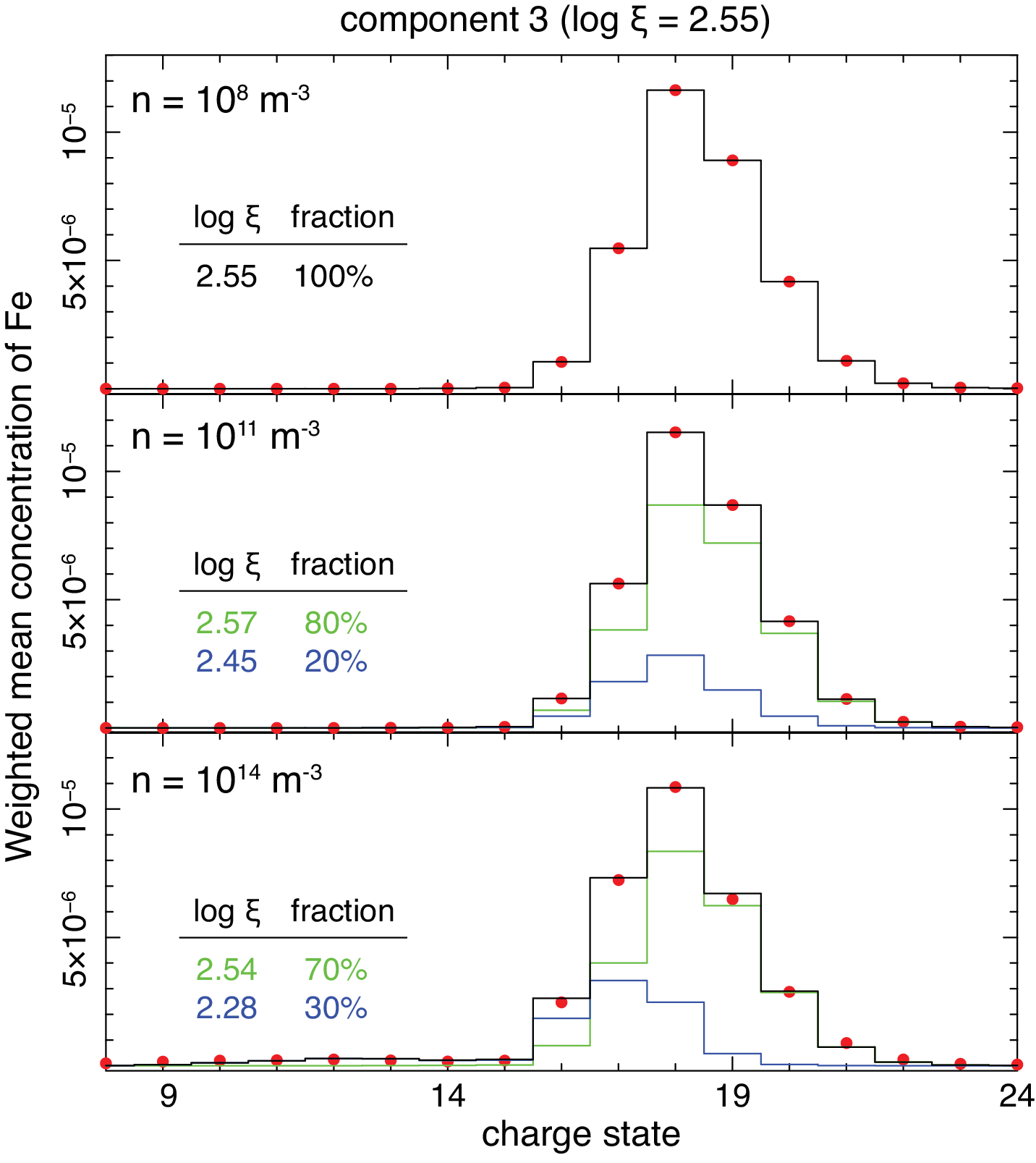}}
\caption{The average concentration distributions of Fe ions obtained from simulation runs with gas densities of $10^{8}$, $10^{11}$, and $10^{14}$ m$^{-3}$ for component 3. The obtained ion concentrations are illustrated as red dots, and the best-fit total concentration is represented by black lines using two equilibrium components depicted by blue and green lines. The ionization parameters and the fractional contributions of the two equilibrium components are also presented in the plot. A single component is sufficient to fit the average concentration in the case of $10^{8}$ m$^{-3}$. }
\label{fig:icon1}
\end{figure}

\subsection{Effects on ion concentration}
\label{sec:1.2}

\begin{figure}[!htbp]
\centering
\resizebox{1.0\hsize}{!}{\includegraphics[angle=0]{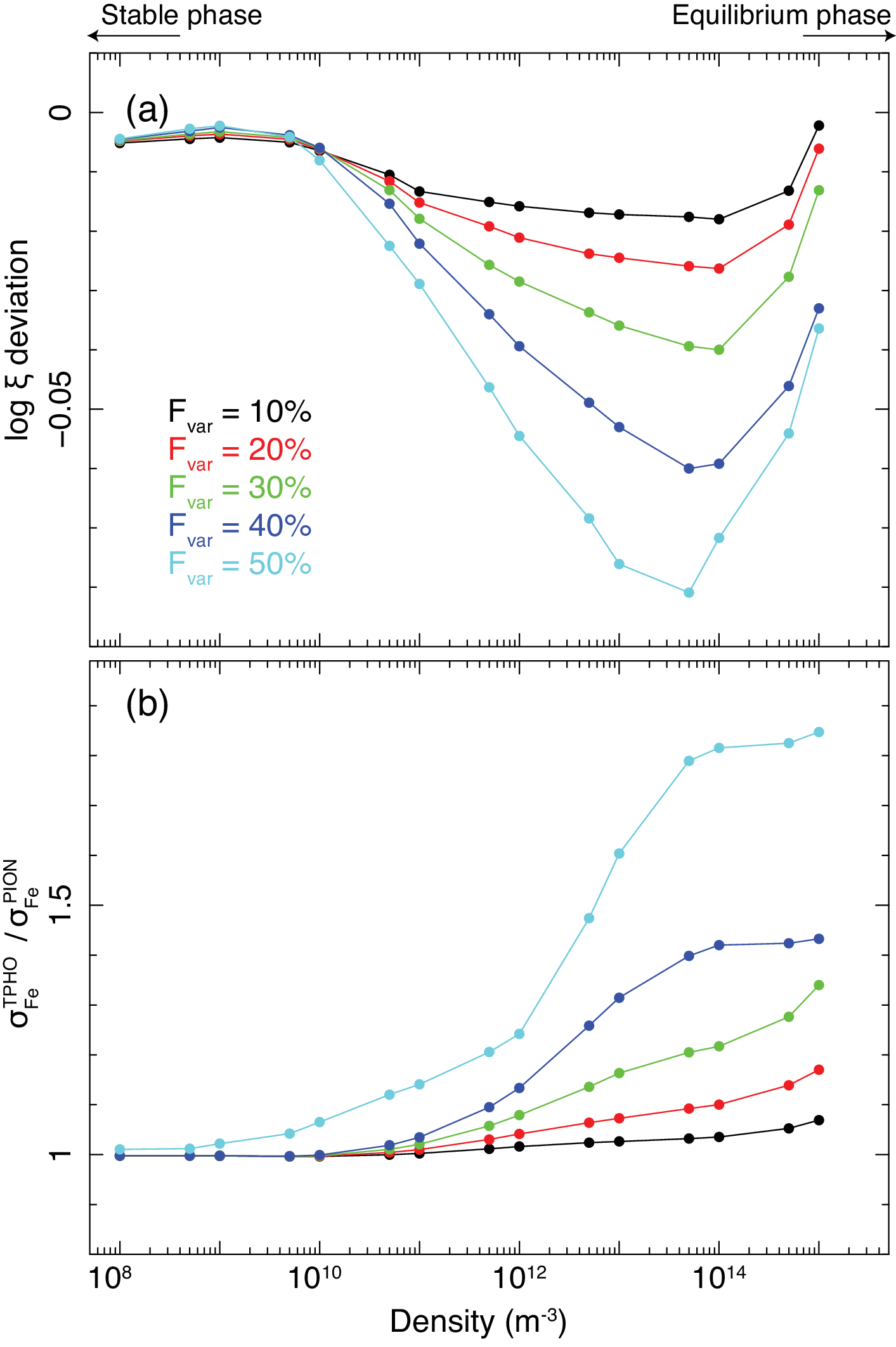}}
\caption{Deviation of log $\xi$ (a) and the dispersion of the Fe ion concentration (b) are plotted as a function of gas densities for lightcurves with fractional variability ranging from 10\% to 50\% (the simulated lightcurve shown in Figure~\ref{fig:lclong} has an average fractional variability $F_{\rm var}$ of $\sim 40$\%). The deviation of log $\xi$ is calculated as the difference between the intrinsic value and the value obtained from a single equilibrium component fitting of the simulated Fe ion concentration distribution such as the ones shown in Figure~\ref{fig:icon1}. The curves representing the averages over the nine warm absorber components (see Table~\ref{tab_pion}) are plotted for both the deviation and dispersion.   }
\label{fig:dlogxi}
\end{figure}


To characterize the effect on ion concentration for a time-averaged absorption spectrum, we perform a full {\tt tpho} run based on a simulated lightcurve, which enables us to track the time-dependent evolution of ion concentration throughout the period of interest. A weighting factor based on the square root of the X-ray flux is applied to determine the mean ion concentration for a time-integrated spectrum. The gas densities of the nine warm absorber components are set to vary between 10$^8$ m$^{-3}$ and 10$^{15}$ m$^{-3}$. To make sure that the results are statistically robust, each simulation is conducted for a considerable duration of $\sim 5 \times 10^{7}$ seconds as shown in Figure~\ref{fig:lclong}. With the {\tt tpho} model it is assumed that the ionizing SED retains its shape over the period.

First we examine how the ionization state of the individual warm absorber component changes over time. Figure~\ref{fig:lclong} provides a visual representation of the temporal evolution of the Fe average charge of component 3 (Table~\ref{tab_pion}) in the warm absorber. The average charge $C_{\rm Fe}$ is computed by $C_{\rm Fe} = \Sigma^{25}_{i=0} c({\rm Fe}^{i+})\times i / \Sigma^{25}_{i=0} c({\rm Fe}^{i+})$, where $c({\rm Fe}^{i+})$ represents the concentration of Fe ions with a charge $i$. Component 3 is chosen as it encompasses a wide span of Fe ion species given its log $\xi$ of 2.55 \citep{m2019}, providing a clear visualization of the potential changes. The average charge exhibits a minor delay of $\leq 10^{4}$ seconds with respect to the lightcurve for a gas density of around $10^{14}$ m$^{-3}$. As the gas density decreases, the amplitude of the charge variation decreases while the delay between the lightcurve and the average charge evolution increases. The underlying density-delay relation presented in this figure is consistent with those reported in previous studies on time-dependent evolution of warm absorber states \citep{j2022, r2022, l2023}.

In Figure~\ref{fig:icon1}, we plot the mean ion concentrations of Fe with three different gas densities for the warm absorber component 3. At low density, the ion concentration can be well reproduced by a single equilibrium {\tt pion} component, indicating a stable state where the effect on ion concentration by a changing SED is minimal. While for a high density cloud, the obtained mean ion concentration needs to be approximated by two equilibrium components with distinct ionization parameters. Furthermore, the normalizations of the two equilibrium components tend to converge at high densities. This behavior arises from two factors. Firstly, for a fixed ionization parameter and SED, the product of density and distance from the source remains constant. As the density increases, the distance from the source decreases, leading to a higher ionizing flux. At the same time, the rate of recombination due to collisions scales with density in the same way. These factors together result in a larger fluctuation in the ion concentration and a broader mean charge distribution at high densities.

Here we conduct a systematic comparison between time-evolving and equilibrium spectra. For each warm absorber component, a single {\tt pion} distribution is utilized to fit the simulated {\tt tpho} ion concentration of Fe. This fitting process allows us to determine the deviation of log~$\xi$ from its original value, e.g., log~$\xi$ = 2.55 for component 3. To quantify the overall effect, we calculate the average log $\xi$ deviations by taking the mean across all nine components. Note that this approach would however overlook the potential differences between the high and low ionization components (e.g., \citealt{j2022}). As shown in Figure~\ref{fig:dlogxi}, the log $\xi$ deviations is negligible at low densities when the cloud maintains an approximately ``stable'' state throughout the simulation with a recombination timescale of $\sim 10^{10}$ seconds, which is longer than the entire simulation duration. However, as the density increases, the {\tt pion}-measured log $\xi$ tends to underestimate the intrinsic value by approximately 0.05 at a density of $10^{13} - 10^{14}$ m$^{-3}$, given a typical lightcurve fractional variability of $\sim 40$\%. The deviation of log $\xi$ becomes small again once the density reaches $10^{15}$ m$^{-3}$, as the cloud recombination timescale decreases to $\leq 10^{3}$ seconds, making it considerably shorter than the lightcurve variability timescale. On the other hand, at high densities, the oscillation of the ionization state results in a more even distribution of charge states. Figure~\ref{fig:dlogxi} demonstrates that for a cloud density of $10^{14}$ m$^{-3}$, the dispersion in ion concentration $\sigma_{\rm Fe}$ is 40\% higher than that predicted by the equilibrium solution, assuming a lightcurve fractional variability of approximately 40\%. Therefore, it can be inferred that the ionization parameter may be underestimated by the equilibrium solution in situations where the plasma is in a ``delayed'' state \citep{r2022, l2023} where recombination takes place at a timescale that is comparable to the lightcurve variability. In situations where the timescale of recombination is much longer  (``stable'') or shorter (``equilibrium'') than the timescale of source variation, the central ionization parameters obtained in equilibrium would be less biased, although the ion concentration becomes significantly more dispersed than a single {\tt pion} component in the ``equilibrium'' scenario as also illustrated in Figure~\ref{fig:icon1}. 

\begin{figure}[!htbp]
\centering
\resizebox{1.0\hsize}{!}{\includegraphics[angle=0]{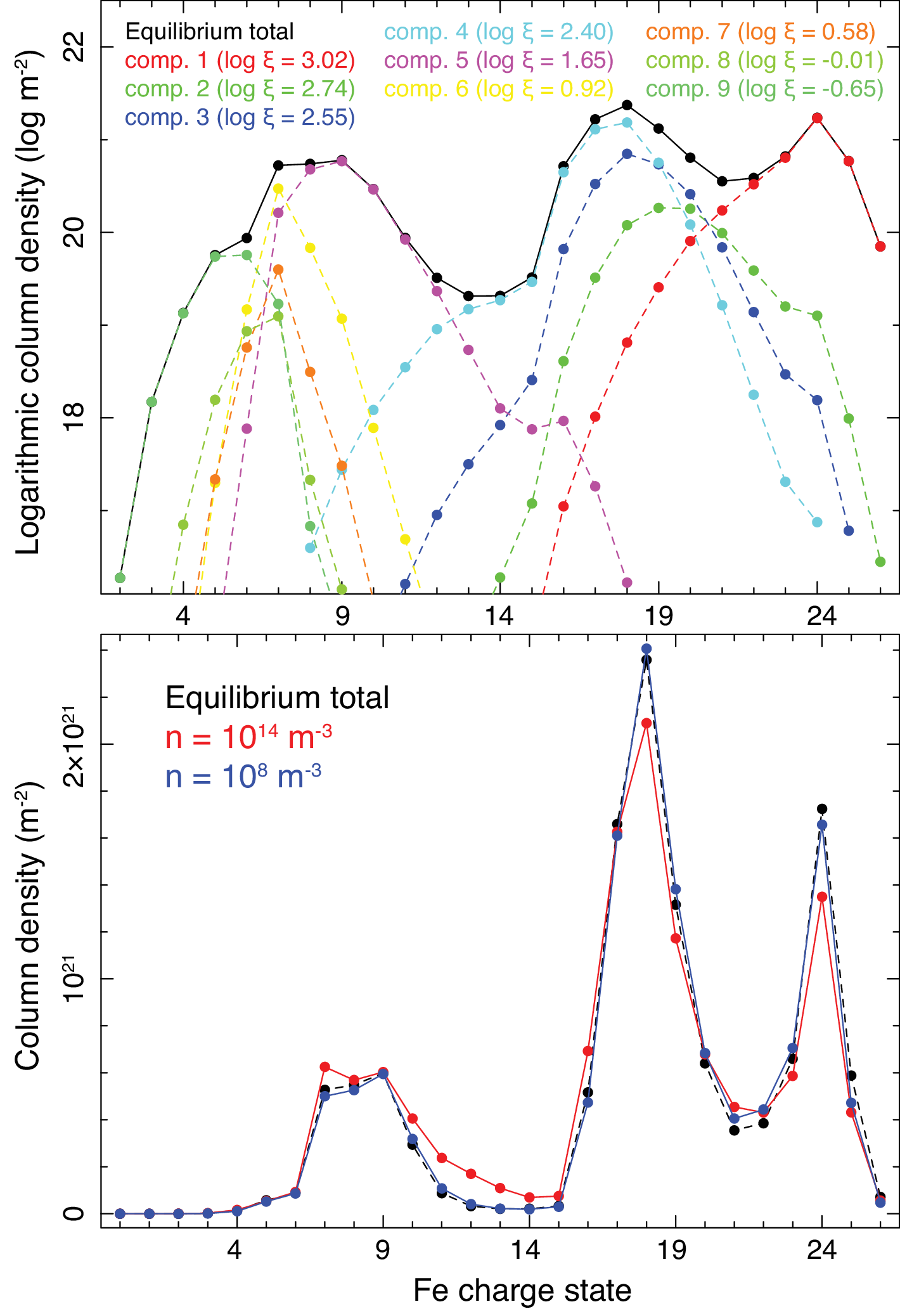}}
\caption{Column densities of Fe in equilibrium and the time-evolving condition. The upper panel of the figure shows the total column density of Fe in equilibrium, with each individual component plotted in a different color. To capture the wide range of concentrations spanning several orders of magnitude, a logarithmic scale is employed. The lower panel compares the equilibrium column density (represented by a black dashed line) to the average column density of time-dependent photoionization with gas densities of $10^{8}$ and $10^{14}$ m$^{-3}$ shown in blue and red lines.  }
\label{fig:piontpho}
\end{figure}

In Figure~\ref{fig:piontpho}, we present the accumulated ionic column densities of Fe as a function of gas density throughout the simulated period. These column densities are obtained by summing the contributions from the nine warm absorber components in NGC~3783. In this calculation, all components are assumed to have the same density. The results reveal that, for a gas density of $10^{8}$ m$^{-3}$, the ionic column density closely aligns with the {\tt pion} model assuming ionization and thermal equilibrium. However, at a higher gas density of $10^{14}$ m$^{-3}$, the curve deviates significantly from the {\tt pion} solution, particularly at the peaks corresponding to \ion{Fe}{XIX} and \ion{Fe}{XXV}. Moreover, the {\tt tpho} curve exhibits a slightly larger concentration of low ionization species (\ion{Fe}{VIII} to \ion{Fe}{XV}) compared to the {\tt pion} case. This discrepancy can be, at least partly, attributed to the dispersion effect in ion concentration (as shown in Figure~\ref{fig:dlogxi}b) caused by the decreasing recombination timescale. These results demonstrate the potential of absorption spectroscopy using a proper time-dependent model to offer useful constraints on gas density.

\section{Application to the 2000 and 2001 Chandra/HETGS observations}
\label{sec:app}

\begin{table}[!htbp]
    \centering
    \caption{NGC~3783 {\it Chandra} HETGS data}    
    \begin{tabular}{lcc}
       \hline
       \hline
        ID & Start Time & Exposure \\
           &            & (ks)     \\
         \hline
        373   & 2000-01-20 & 56      \\
        2090  & 2001-02-24 & 166     \\
        2091  & 2001-02-27 & 169     \\
        2092  & 2001-03-10 & 165     \\
        2093  & 2001-03-31 & 166     \\
        2094  & 2001-06-26 & 166     \\
        \hline
        \hline
    \end{tabular}
    \label{obs_log}
\end{table}

\begin{figure}[!htbp]
\centering
\resizebox{1.0\hsize}{!}{\includegraphics[angle=0]{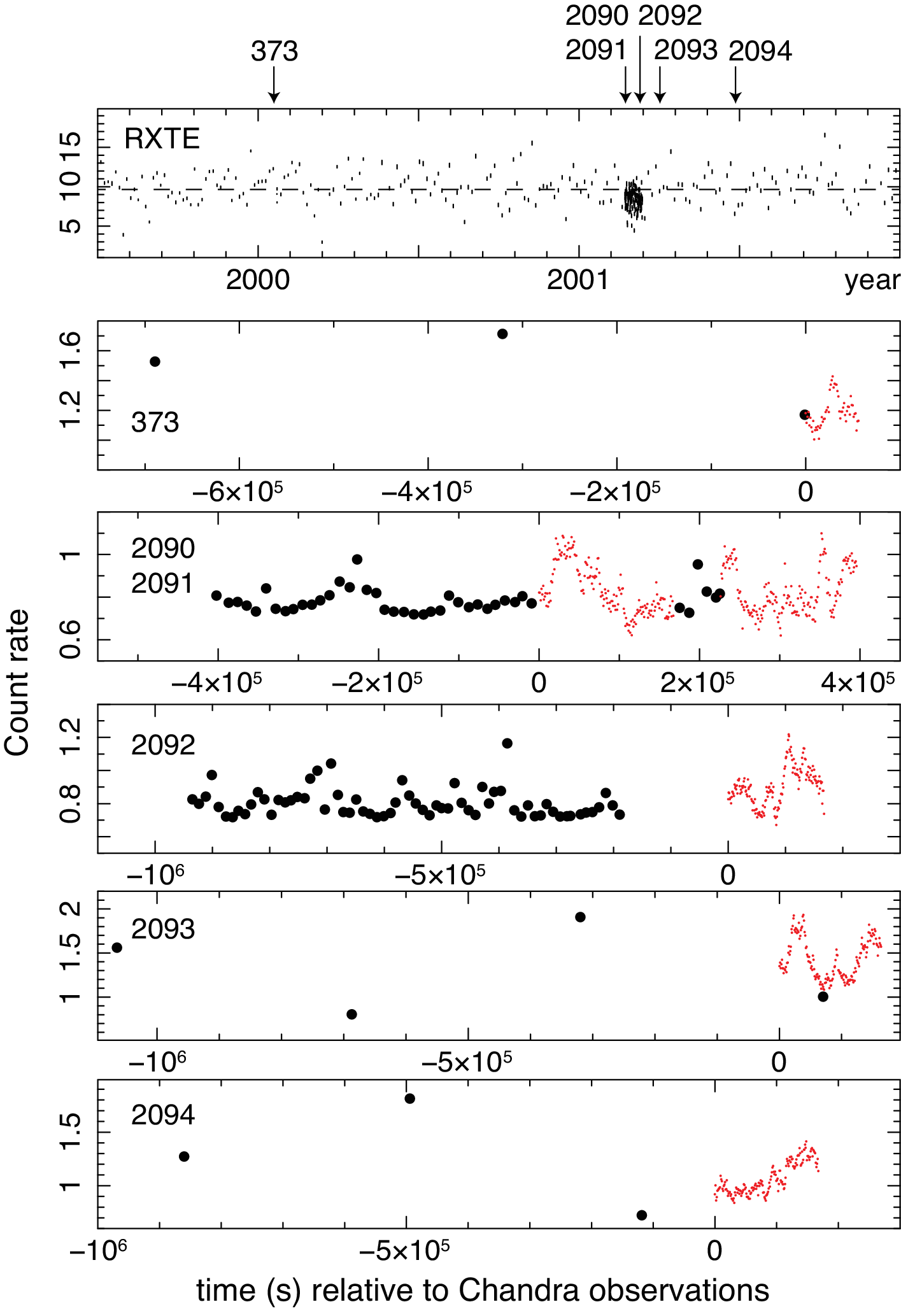}}
\caption{{\it RXTE} and {\it Chandra} HETGS lightcurves of NGC~3783 from mid 1999 to the end of 2001. The upper panel shows the {\it RXTE} proportional counter array (PCA) data in the $2-60$~keV band, with the average flux level indicated by a horizontal line. The {\it Chandra} HETGS lightcurves are presented in the panels below as red data points, along with the adjacent {\it RXTE} data converted using the {\it PIMMS} tool. }
\label{fig:lc}
\end{figure}

\begin{figure}[!htbp]
\centering
\resizebox{1.0\hsize}{!}{\includegraphics[angle=0]{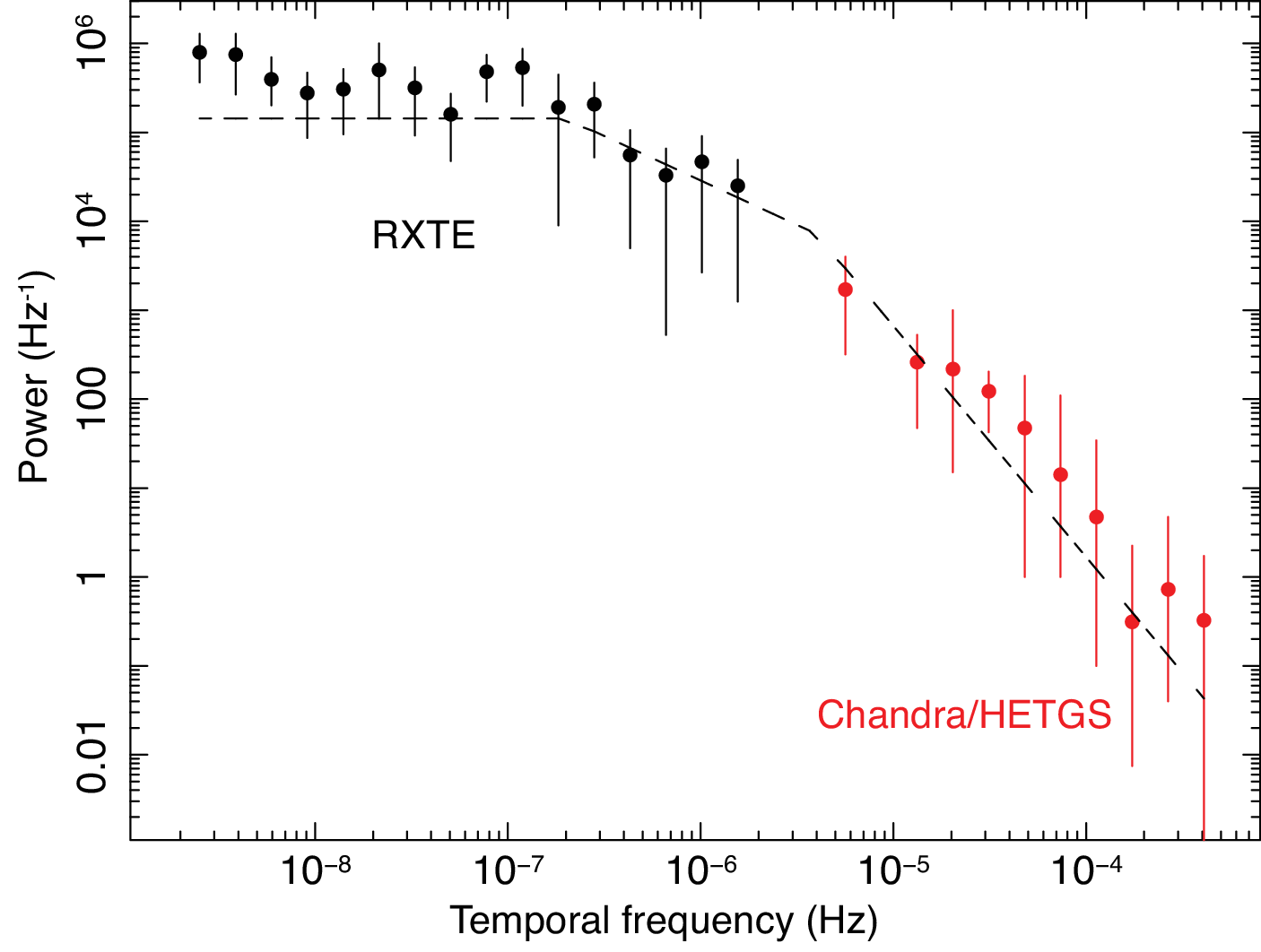}}
\caption{Power spectral density function of the combined {\it RXTE} (black) and {\it Chandra} HETGS (red) lightcurves shown in Figure~\ref{fig:lc}. The data are compared with the model (dashed line) based on the best-fit results obtained in \citet{m2005} on the data in the $0.2-12.0$~keV range. }
\label{fig:ps}
\end{figure}

\begin{figure*}[!htbp]
\centering
\resizebox{0.85\hsize}{!}{\includegraphics[angle=0]{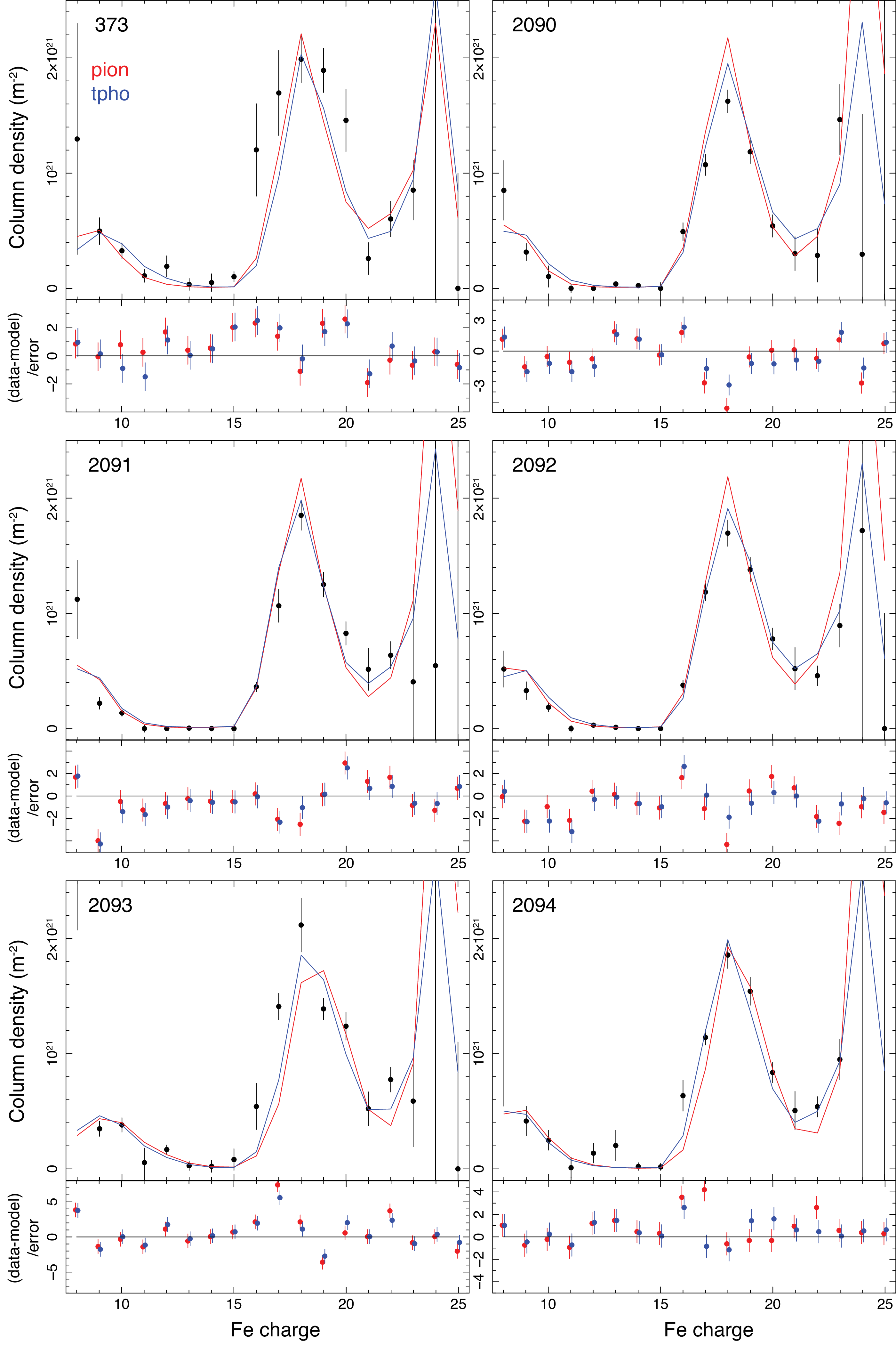}}
\caption{Column densities of all Fe ions ranging from \ion{Fe}{IX} to \ion{Fe}{XXVI} determined through fitting using the {\tt slab} model (black data points). The observed data points are compared with the values computed using the {\tt pion} (red) and {\tt tpho} (blue) models. Each panel in the figure represents one HETGS observation.}
\label{fig:concom}
\end{figure*}

\begin{figure*}[!htbp]
\centering
\resizebox{0.9\hsize}{!}{\includegraphics[angle=0]{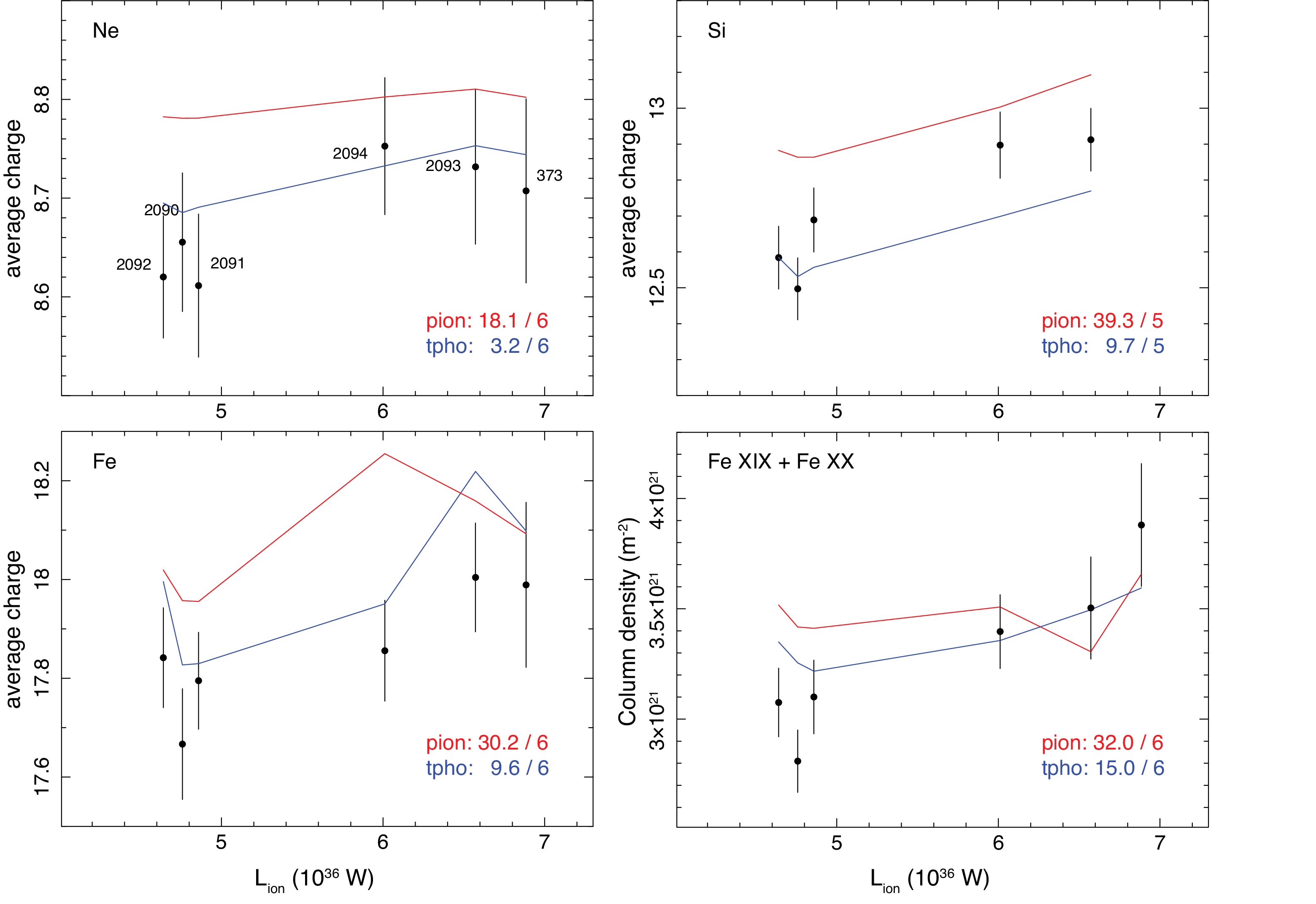}}
\caption{The average charge states of Ne, Si, and Fe, as well as the combined column densities of \ion{Fe}{XIX} and \ion{Fe}{XX}, are plotted as a function of the ionizing luminosity. The average charges are defined in \S~\ref{sec:1.2}. Each data point corresponds to an individual observation and was obtained by fitting with the {\tt slab} model. The values obtained from the {\tt pion} and {\tt tpho} models are represented by the red and blue curves, respectively. The chi-square values between the models and the data are displayed as numbers in the lower right corner.}
\label{fig:xlfe}
\end{figure*}

In order to investigate the feasibility of our approach on actual observations, we perform a self-consistent {\tt tpho} modeling on the {\it Chandra} high energy transmission grating spectrometer (HETGS) data obtained during the unobscured period of 2000-2001 (Table~\ref{obs_log}). We limit our analysis to a relatively small set of data as the current {\tt tpho} model still requires a significant amount of computational resource which prevents us from extending our analysis to data of other time periods or instruments. To process the data, we utilize the CIAO v4.15 software and calibration database (CALDB) v4.10. The {\tt chandra\_repro} script is used to perform data screening and generate spectral files for each exposure. We then merge the spectra for orders $\pm1$ using the CIAO {\tt combine\_grating\_spectra} tool, along with the associated response files. We perform a joint fit on the medium energy grating (MEG) spectrum covering the wavelength range of $6-16.5$~{\AA}, and the high energy grating (HEG) spectrum in the $1.8-3.1$~{\AA} range.

Figure~\ref{fig:lc} depicts the {\it RXTE} and HETGS lightcurves of NGC~3783 from July 1999 to December 2001.
To perform the time-dependent calculation for each {\it Chandra} observation, we utilize the HETGS lightcurve in combination with the {\it RXTE} lightcurve obtained prior to the {\it Chandra} observation. By incorporating the {\it RXTE} lightcurve, we are able to track the changes in the source and model the possible delayed response in the ionization state which can potentially have an impact on the HETGS data. The {\it PIMMS} tool is used to convert the {\it RXTE} count rate to the HETGS rate based on the broad-band spectral model described in \S~\ref{sec:3.1}. As shown in Figure~\ref{fig:lc}, during a few overlapping periods of the two instruments, the converted RXTE count rates match well with the HETGS rates. Due to the limited coverage of the {\it RXTE} data for observations 1, 5, and 6, the short-term variability is not well captured, possibly introducing a systematic bias. Further details regarding this uncertainty will be discussed in Section~\ref{3.2}.

Using the method outlined in \citet{u2002} and \citet{m2005}, we have computed the combined power spectrum for the {\it RXTE} and {\it Chandra} data taken between July 1999 and December 2001. Figure~\ref{fig:ps} shows that the resulting spectrum agrees well with the doubly broken model presented in \citet{m2005}, which was also utilized to generate the lightcurve used in \S~\ref{sec:1}. This agreement indicates that the variability observed during the {\it Chandra} observations aligns well with the average properties of this AGN. Considering that the primary variability of the source arises from variations detected by {\it RXTE} below the break frequency of $4 \times 10^{-6}$~Hz, the {\it Chandra} exposures would capture mostly the tail end of the variability on shorter timescales.

\subsection{Spectral analysis}
\label{sec:3.1}

For each HETGS exposure, we fit one time-integrated spectrum by utilizing a time-evolving scheme as follows. By adopting the approach presented in \citet{m2017}, \citet{k2018}, and \citet{m2019}, we first build up a contemporaneous SED which serves as the photoionization continuum for the {\tt pion} and {\tt tpho} components. This continuum model includes a Comptonized disk component ({\tt comt}), a power-law with cut-off ({\tt pow}), and a neutral reflection component ({\tt refl}). Together, these components form the intrinsic SED. We allow the normalizations of the three continuum components to vary between different pointings to account for the intrinsic variability of the nucleus as indicated in the lightcurve. To model the warm absorbers, we use a total of nine quiescent {\tt pion} absorption components, where the column densities, ionization parameters, outflow, and microscopic turbulence velocities are fixed to those obtained in \citet{m2019}. For the photoionized emission, we use three {\tt pion} components with positive covering relative to the nucleus, two for the X-ray narrow emission feature and one for the broad features. The properties of these emission {\tt pion} components are also set to the values determined in \citet{m2019}. By fitting the global spectrum, we determine the contemporaneous SED for each HETGS pointing. 

We further explore the modeling of the warm absorber components using the {\tt pion} model. For warm absorber components 1 to 5, we allow the column densities and ionization parameters to vary, while keeping the outflow and microscopic turbulence velocities consistent with the values from \citet{m2019}. To account for variations in the ionization parameter across different observations, we assume that the hydrogen densities and distances of the warm absorbers remain constant. As a result, the ionization parameter for each observation is scaled by the contemporaneous SED obtained above in the $1-10^{3}$ Ryd range. Due to the limited constraining power of the HETGS data on components at low ionization states, we choose to fix the column densities and ionization parameters of components 6 to 9 to the values reported in \citet{m2019}. We assume that each {\tt pion} component extends significantly along the line of sight, with a covering factor of 1. The results obtained from the {\tt pion} fit can be found in Table~\ref{tab_pion}.

We proceed by replacing the nine {\tt pion} absorption components with corresponding {\tt tpho} components. These {\tt tpho} components calculate the time-dependent spectrum based on the obtained lightcurves and contemporaneous SEDs. In the {\tt tpho} modeling, both the {\it Chandra} and adjacent {\it RXTE} lightcurves (Fig.~\ref{fig:lc}) are utilized, while the output model represents the integrated spectrum within the {\it Chandra} period. Therefore, the fitting results provide an average characterization of the properties observed during the {\it Chandra} observations. The basic setup of the {\tt tpho} components remains identical to that of the {\tt pion} components, except for one key difference: the hydrogen number densities are allowed to vary freely in the {\tt tpho} modeling of warm absorber components 1 to 5. This variation in hydrogen number densities enables the modeling of time-dependent behavior within these components, contrasting with the {\tt pion} modeling where the densities are not constrained. For the low ionization components 6 to 9, the column densities and ionization parameters are fixed to the values reported in \citet{m2019}. Their hydrogen number densities are set to their lower limits, effectively turning off their time evolution. The {\tt tpho} results are summarized in Table~\ref{cons_table}.

\begin{table}[!htbp]
    \caption{{\tt pion} constraints on the warm absorber properties}    
    \begin{tabular}{lcccc}
       \hline
       \hline
        Comp. & log $\xi$        & $N_{\rm H}$        & $v_{\rm out}$   & $v_{\rm mic}$  \\
              & 10$^{-9}$ W m    & 10$^{25}$ m$^{-2}$ & km s$^{-1}$     & km s$^{-1}$   \\
         \hline
        1     & 2.98 $\pm$ 0.01  & 13.3 $\pm$ 0.9     & [-480]          & [120]         \\
        2     & 2.65 $\pm$ 0.01  & 1.9 $\pm$ 0.2      & [-1300]         & [120]         \\
        3     & 2.40 $\pm$ 0.01  & 7.9 $\pm$ 0.7      & [-830]          & [46]          \\
        4     & 2.38 $\pm$ 0.01  & 7.9 $\pm$ 0.4      & [-460]          & [46]          \\
        5     & 1.63 $\pm$ 0.01  & 4.5 $\pm$ 0.3      & [-575]          & [46]          \\
        6     & [0.92]           & [1.2]              & [-1170]         & [46]          \\
        7     & [0.58]           & [0.15]             & [-1070]         & [46]          \\
        8     & [-0.01]          & [0.07]             & [-1600]         & [790]         \\
        9     & [-0.65]          & [0.44]             & [-1100]         & [790]         \\
        
        \hline
        \hline
    \end{tabular}
    Note: Ionization parameters and column densities obtained from fitting with the {\tt pion} model.
    The ionization parameters represent the mean values of the six observations. There are minor differences between our values and those presented by \citet{m2019} on components $1-4$, potentially due to the fact that \citet{m2019} utilized both {\it Chandra} and {\it XMM-Newton} data in the analysis.
Values enclosed in square brackets represent fixed parameters that are set to the results reported in \citet{m2019} using time-integrated spectra. \\
    \label{tab_pion}
\end{table}

The fit utilizing {\tt tpho} component results in a C-statistic of 12116 for an expected value of 8073, which is a mild improvement compared to the C-statistic of 12889 obtained with the {\tt pion} model. In order to gain insight into the enhanced modeling of the spectral features by {\tt tpho}, we put forward an analysis employing the {\tt slab} model. This model allows us to independently determine the column density of each ionic species, without relying on assumptions regarding the ionization balance. Thereby the {\tt slab} component can be considered as a complete, independent representation of the ionization state of the warm absorbers. To mitigate the level of degeneracy, we employ two {\tt slab} components, instead of nine, to model the absorption feature for each HETGS observation. The first component is introduced with a velocity varying in the range from -400 to -600 km s$^{-1}$, representing warm absorber components 1, 4, and 5. The second {\tt slab} component encompasses an outflow velocity ranging from -800 to -1600 km s$^{-1}$ and combines warm absorber components 2, 3, 6, 7, 8, and 9. The turbulence velocities of the two {\tt slab} components are fitted freely. The fit utilizing the two {\tt slab} components yields a satisfactory C-statistic of 11844 for an expected value of 8058. The addition of a third {\tt slab} component does not yield any further significant improvement in the fitting quality.

By substituting the nine {\tt tpho} components with these two {\tt slab} components, we fit the set of HETGS data again.
We permit the column densities of Fe species from \ion{Fe}{IX} to \ion{Fe}{XXVI}, Ne ions from \ion{Ne}{III} to \ion{Ne}{X}, Mg ions from \ion{Mg}{VIII} to \ion{Mg}{XII}, Si ions from \ion{Si}{X} to \ion{Si}{XIV}, and S ions from \ion{S}{XI} to \ion{S}{XVI} to vary freely for both $\tt slab$ components. To reduce degeneracy, we keep the column densities of O, Na, Al, Ar, Ca, Ni, as well as the remaining species of Ne, Mg, Si, S, and Fe, at the fixed values determined from the global {\tt tpho} fit. The column densities combining the two $\tt slab$ components can be thus used as the absolute absorption values for each observation.

Figure~\ref{fig:concom} displays a comparison of the ionic column densities ranging from \ion{Fe}{IX} to \ion{Fe}{XXVI} obtained from three different models: {\tt pion}, {\tt tpho}, and {\tt slab}, for each HETGS pointing. All datasets exhibit two prominent peaks centered around \ion{Fe}{XIX} and \ion{Fe}{XXV}, in line with the average model depicted in Figure~\ref{fig:piontpho}. While the current HETGS data lack sufficient constraints on the \ion{Fe}{XXV} peak, it becomes evident that the {\tt tpho} model systematically provides a better or, at the very least, comparable fit to the \ion{Fe}{XIX} peak. The {\tt pion} curves tend to slightly overestimate the peak values in observations 2090, 2091, and 2092. For observation 2093, the {\tt pion} model seems to underestimate the column density of \ion{Fe}{XIX}, meanwhile overestimating that of \ion{Fe}{XX}. Some of these features can be better reproduced by utilizing the {\tt tpho} components. Both models perform similarly when compared to the {\tt slab} results for observations 373 and 2094.

To further compare the results between the {\tt slab} and the physical models, we compute the average charges of Ne to Fe using the $\tt slab$ fits for each observation. The average charges are then plotted in Figure~\ref{fig:xlfe} as a function of the contemporaneous ionizing fluxes, alongside the average charges calculated using the best-fit $\tt pion$ and $\tt tpho$ models. Due to limited constraints, the Mg and S results are not included in the comparison. As the ionizing fluxes increase, we observe a corresponding mild increase in the obtained average charges, indicating for evolving ionization levels under varying ionizing fluxes within a finite recombination time. A similar pattern is found in the total column densities of \ion{Fe}{XIX} and \ion{Fe}{XX}, which are the predominant iron species. Conducting a statistical comparison among the {\tt slab}, $\tt pion$, and $\tt tpho$ results reveals that the $\tt tpho$ model consistently aligns better with the {\tt slab} results. The $\tt pion$ calculations appear to slightly overestimate the average charges at low ionizing luminosity. This is basically in line with the results depicted in Figure~\ref{fig:concom}.

Both the {\tt pion} and {\tt tpho} models utilize identical atomic data for calculating ionization balance, thermal balance, and absorption lines. The key distinction lies in the fact that the {\tt tpho} model introduces time and density dependencies in the ionization and thermal states, whereas the {\tt pion} model adopts equilibrium for both. The better agreement between {\tt tpho} and {\tt slab} results thus provides evidence for non-equilibrium states within each observation. This further indicates the presence of non-zero densities for the relevant warm absorption components.

\subsection{Systematic uncertainties}
\label{3.2}
An apparent concern with the current {\tt tpho} approach arises from the combination of $\it RXTE$ and $\it Chandra$ light curves, particularly where the $\it RXTE$ light curves are sparsely sampled in observations 373, 2093, and 2094. Consequently, the $\it RXTE$ data may not adequately capture the full history of source variations preceding the {\it Chandra} observation, potentially introducing systematic biases on the ion concentration variations during the target periods due to the finite recombination time. To estimate the magnitude of this uncertainty, we explore an extreme scenario where each $\it RXTE$ light curve is transformed into a flat constant at its average value. Consequently, the ion concentration remains constant throughout each $\it RXTE$ period, resulting in minimal impact on the subsequent variations. By fitting the HETGS data using these modified lightcurves, we identify fractional differences of $\sim 5-15$\% in the derived densities of the warm absorber components when compared to the values obtained using the original lightcurves. Therefore, it is unlikely that the uncertainty in $\it RXTE$ lightcurve has substantially affected the current {\tt tpho} results.

\begin{figure}[!htbp]
\centering
\resizebox{1.0\hsize}{!}{\includegraphics[angle=0]{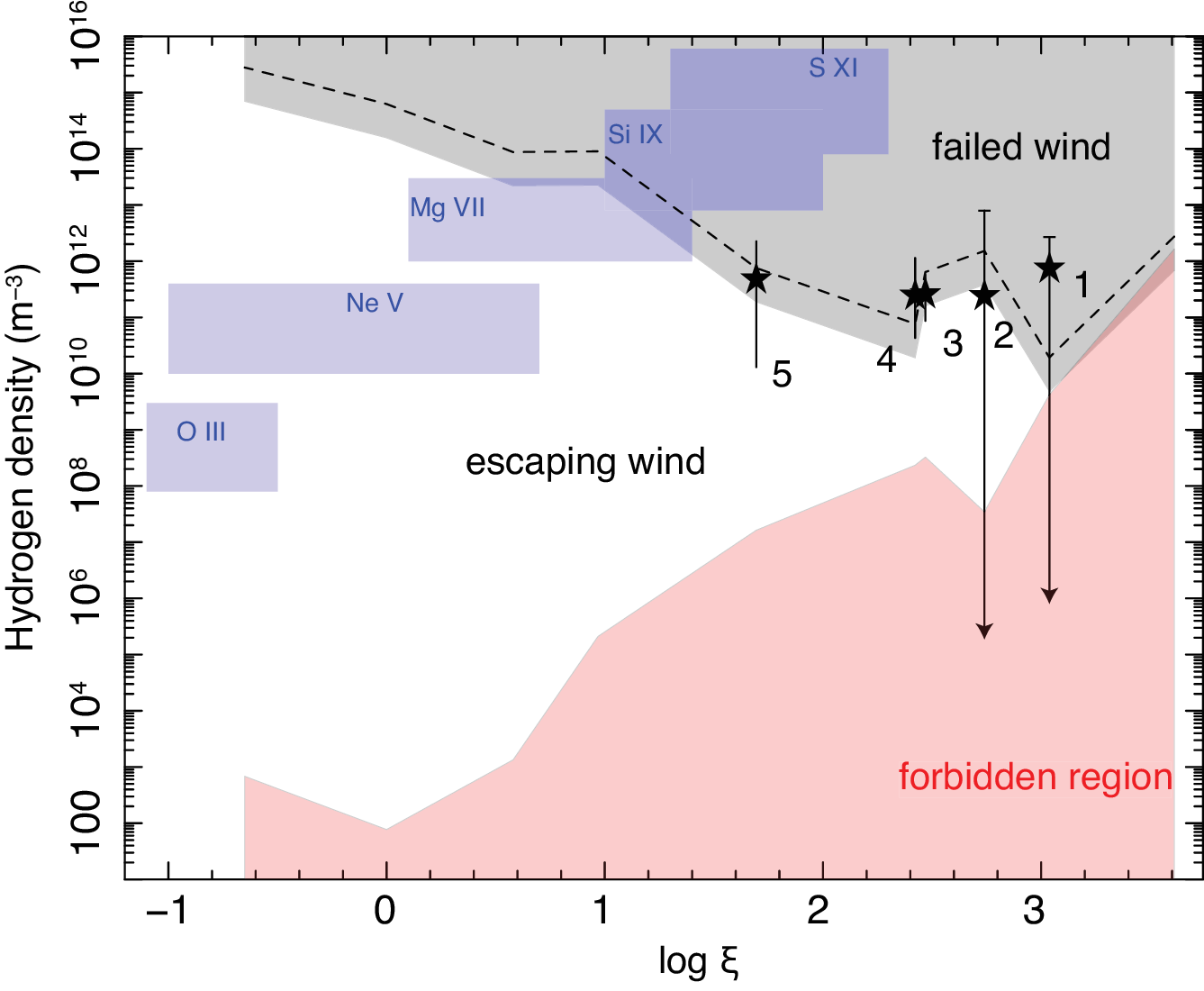}}
\caption{Hydrogen number densities of the warm absorbers plotted as a function of their ionization parameters. Our results for components 1-5 from the {\tt tpho} model are shown as stars. The gray shaded region represents the region where the outflow is unable to escape the gravitational potential of the black hole due to $n_{\rm H} > n_{\rm esc}$, while the dashed line indicates the position of $n^{\ast}_{\rm esc}$ (see \S~\ref{4.4}). The red shaded region shows the forbidden regime where the thickness of the outflow exceeds its distance ($n_{\rm H} < n_{\rm low}$). In addition, the blue shaded regions indicate the areas where density-sensitive X-ray lines can be utilized to measure densities.}
\label{fig:hd_v}
\end{figure}

In addition, the assumed equilibrium solution for the initial ionization states of the warm absorber components, obtained from a time-integrated analysis, might introduce bias, as the true condition is likely to be non-equilibrium. Determining the precise initial ionization state presents a challenge in itself. However, we can assess the uncertainty associated with the initial condition by modifying the first data point in the lightcurve. This would have a significant impact on the evolution of the ionization states since the ionizing flux utilized in the {\tt tpho} model is a relative value scaled by the first data point in the lightcurve. To evaluate this uncertainty, we alter the first data point by a fractional change of 15\%, determined from the standard deviation of the HETGS lightcurves averaged over six observations. The modified lightcurve results in a fractional difference of $\sim 40-80$\% in the best-fit densities. This suggests that the current assumption regarding the initial ionization state might introduce a substantial systematic uncertainty into the results. The reported density errors in Table~\ref{cons_table} and Figure~\ref{fig:hd_v} are derived by combining the statistical errors with the systematic errors stemming from both the $\it RXTE$ lightcurve and the initial state.

\section{Interpretation and future prospect}
\label{sec:3}

\begin{table*}[!htbp]
    \caption{{\tt tpho} and UV constraints on the physical properties of the warm absorbers}    
    \begin{tabular}{lcccccccc}
       \hline
       \hline
        Comp.     & log $\xi^{a}$       &   $N_{\rm H}$     &    $n_{\rm H}$      &  $P^{b}$               &  $t_{\rm rec}^{c}$          &   $R^{d}$            &   $\Delta R^{e}$    \\
                  & 10$^{-9}$ W m    &10$^{25}$ m$^{-2}$ & 10$^{11}$ m$^{-3}$  &  10$^{10}$ keV m$^{-3}$&  10$^{4}$ s           & 10$^{15}$ m        &   10$^{14}$ m     \\
         \hline
                1 & 3.04 $\pm$ 0.01  & 15.7 $\pm$ 1.0   &    $\leq 23.1$      &  $\leq 104.3$           &  $\geq 0.6$           & $\geq 1.6$         &  $\geq 0.7$       \\
                2 & 2.74 $\pm$ 0.03  & 2.0  $\pm$ 0.2   &    $\leq 82.4$      &  $\leq 158.0$           &  $\geq 0.2$           & $\geq 1.2$         &  $\geq 0.02$      \\
                3 & 2.47 $\pm$ 0.01  & 8.4  $\pm$ 0.7   &    2.6 $\pm$ 1.7    &  1.5 $\pm$ 1.0          &  5.0$^{+9.7}_{-2.0}$  & 9.1 $\pm$ 2.3      &   3.2 $\pm$ 1.6   \\
                4 & 2.42 $\pm$ 0.01  & 7.5  $\pm$ 0.3   & 2.4$^{+7.8}_{-1.9}$ &  1.4$^{+4.7}_{-1.1}$    &  5.3$^{+20.8}_{-4.2}$ &10.0$^{+2.8}_{-3.0}$&3.1$^{+1.5}_{-1.3}$\\
                5 & 1.69 $\pm$ 0.01  & 4.6  $\pm$ 0.3   & 4.7$^{+14.3}_{-4.5}$&  0.5$^{+1.4}_{-0.4}$    & 0.4$^{+9.3}_{-0.3}$  &16.6$^{+5.8}_{-5.0}$&1.0$^{+0.8}_{-0.4}$\\
             UV1a$^{f}$ & -0.47      & 0.4              &    0.3              &  0.004                  &  139.3                & 871.0              & 1.6 \\
             UV1b & 0.75             & 0.2$-$2.5        &    0.02             &  0.0006                 &  347.5                & 871.0              & 12.6$-$158.5 \\
             UV2  & 0.69             & 0.3              &    $\geq$ 0.02      &  $\geq$ 0.0005          &  $\leq 388.8$         & $\leq$ 794.3       & $\leq 10$    \\
             UV3  & 0.64             & 1.3              &    $\geq$ 0.01      &  $\geq$ 0.0003          &  $\leq 853.9$         & $\leq$ 1584.9      & $\leq$ 158.5 \\
        \hline
        \hline
    \end{tabular}
    
    $^a$ Mean ionization parameters of the six observations. \\
    $^b$ Thermal pressure calculated from electron temperature and density obtained with {\tt tpho}. \\
    $^c$ Recombination time of Fe ions calculated based on charge state change \citep{l2023}. \\
    $^d$ Distance to the central AGN obtained by $R = \sqrt{ L / n_{\rm H} \xi}$. \\
    $^e$ Radial depth assuming uniform, constant density outflow. \\
    $^f$ Properties of UV absorbers obtained with the data of Space Telescope Imaging Spectrograph (STIS) onboard the {\it Hubble} {\it Space} {\it Telescope} \citep{gabel2005}. \\
    \label{cons_table}
\end{table*}

\subsection{Comparing the time-dependent and equilibrium models}

The equilibrium model continues to serve as the current baseline in the examination of photoionized sources, despite that some of such sources potentially exhibit substantial deviations from equilibrium. Consequently, it is crucial to address the similarity and discrepancy between time-dependent and equilibrium models in terms of their spectra. As indicated earlier in \S~\ref{sec:1}, the ion concentrations derived from the equilibrium {\tt pion} model and the time-dependent {\tt tpho} model differ in at least two notable aspects: the ionization parameters obtained with {\tt pion} may exhibit a slight deviation from those with {\tt tpho} within the density range of $10^{11} - 10^{15}$ m$^{-3}$ in the case of NGC~3783, and {\tt tpho} tends to yield a broader range of ion species compared to the equilibrial distribution. 

As shown in Table~\ref{tab_pion} and Table~\ref{cons_table}, the ionization parameters derived from the {\it pion} model exhibit systematically lower values compared to those obtained using the {\tt tpho} model, with a difference ranging from $0.05$ to $0.1$ across various warm absorber components. This aligns well with the expectation illustrated in Figure~\ref{fig:dlogxi}. Despite of the small deviation, it is reasonable to take the warm absorber components identified using the classic {\it pion} model in the HETGS band for this object as a valid approximation.

\subsection{Constraints on densities and distances of the X-ray absorbers}

Although the overall spectral fits with the {\tt tpho} and {\tt pion} models yield a marginal difference (C-statistic of 12116 with {\tt tpho} and 12889 with {\tt pion}), the {\tt tpho} model shows significant improvements over the {\tt pion} model in various instances. This is evident in Figure~\ref{fig:xlfe}, where the average charge state distributions of Ne, Si, and Fe obtained with the {\tt slab} significantly favor the {\tt tpho} model over the {\tt pion} model. Figure~\ref{fig:concom} shows that the {\tt tpho} model consistently better reproduces the ionization peak centered around \ion{Fe}{XIX}, primarily originating from warm absorber components 3 and 4 as depicted in Figure~\ref{fig:piontpho}. Consequently, this enables us to derive useful constraints on the cloud densities associated with components 3 and 4, while obtaining a more marginal constraint for component 5 at a lower ionization state. However, it should be noted that the current HETGS data only allows for the upper limits to be determined for the densities of components 1 and 2, since these components involve highly-ionized Fe where the sensitivity of the present instrument becomes poorer. The obtained density constraints are shown in Figure~\ref{fig:hd_v} and Table~\ref{cons_table}. 

By employing the derived hydrogen number densities, ionization parameters, and source luminosity, we can calculate the distance between the absorber components and the central source. Based on observed densities in the range of several times 10$^{11}$ m$^{-3}$, the distances for components 3-5 are obtained to be around $1-2 \times 10^{16}$ m ($0.3-0.6$ pc). Assuming a roughly uniform distribution of each component with approximately unity volume filling inside the cloud, we can also determine the depth of the cloud along the line of sight from the ratio of column density to number density. This yields depths ranging from $1 \times 10^{14}$ m to $5 \times 10^{14}$ m for these three components. As for components 1 and 2, we can only obtain lower limits for their physical distances and depths.

Next we compare our density measurements with previous findings in the literature. In their study, \citet{netzer2003} compared the absorption features observed during low and high luminosity states with HETGS, suggesting an upper limit of $2.5 \times 10^{11}$ m$^{-3}$ for the highly ionized component and approximately $10^{11}$ m$^{-3}$ for the intermediate component. Through a match considering the ionization parameter and column density, their highly ionized components likely correspond to our components 3 and 4, while their intermediate component partially aligns with our component 5. The density values obtained from our {\tt tpho} model are therefore in line with the limits proposed by \citet{netzer2003}. A similar agreement is found when comparing our density values with the upper limits derived from the variability analysis of {\it XMM-Newton} grating spectra by \citet{behar2003}. By using a thermodynamic modeling of the outflowing structure, \citet{chelouche2005} predicted that the hot X-ray absorbers originate from within a distance of 1~pc, which agrees with our measured values. Finally, by conducting a variability analysis of the Fe L- to M-shell unresolved transition array (UTA) with the HETGS data, \citet{krongold2005} reported a lower limit of approximately $10^{10}$ m$^{-3}$ for the absorber density, and an upper limit on its distance of around 6~pc. These values again align well with our spectroscopic measurements using {\tt tpho}.

By utilizing the reverberation mapping, \citet{peterson2004} reported the presence of a broad line region with radii ranging from approximately $4\times 10^{13}$ m (\ion{He}{II}) to about $3 \times 10^{14}$ m (H$\beta$). A more recent measurement by \citet{gravity2021}, employing the GRAVITY instrument installed at the Very Large Telescope Interferometer (VLTI), provides a broad line region size estimate of approximately $4 \times 10^{14}$ m. Comparing them with the {\tt tpho} results suggests that the warm absorber components 1-5 are located outside the boundary of the broad line region. Furthermore, based on K-band observations using the AMBER instrument at VLTI, \citet{weigelt2012} reported the presence of a ring-like torus with a radius of $5\times 10^{15}$ m (or 0.16~pc), which coincides with the distances of components 1 and 2. In addition, the GRAVITY data revealed a [\ion{Ca}{VIII}]-emitting "coronal line region" with a size of around $1.2 \times 10^{16}$ m. This aligns well with the locations of X-ray warm absorber components 3-5. By using the data obtained from the VLT/SINFONI instrument, \citet{gravity2021} further proposed that the coronal line region is likely an outflowing component with an extension up to approximately $3 \times 10^{18}$ m from the AGN. Overall it suggests a possible scenario in which the X-ray warm absorber components 1-5 represent highly ionized gas clouds distributed between the torus and the inner part of the extended coronal line region.

In Table~\ref{cons_table}, we further present the comparison between the X-ray absorbers and the UV counterparts as reported in \citet{gabel2005}. The UV absorbers are classified into three kinematic components, namely components 1, 2, and 3, with outflow velocities of 1350, 550, and 725 km $s^{-1}$, respectively. Component 1 can be further divided into two physically distinct regions, denoted as 1a and 1b, based on detailed differences in covering factors and kinematic structure. Table~\ref{cons_table} reveals differences between the two sets of absorbers, specifically, the gas densities of the lowly ionized UV absorbers are in the range of $10^{9} - 10^{10}$ m$^{-3}$, which are considerably lower than those of the X-ray components. In addition, the UV components are situated at greater distances from the AGN, with an approximate range of $R \sim 20 - 50$ pc, consistent with the typical location of the AGN inner narrow line region. While the radial depth of UV component 1a appears to align with that of the X-ray components, UV component 1b is significantly more extended along the line of sight assuming a uniform absorber distribution. A shared characteristic between the X-ray and UV components is their relatively small global volume filling in the radial direction, indicated by $\Delta R / R \ll 1$. This feature supports the scenario that both types of absorbers are composed of discrete and compact gas clumps.

\subsection{Constraints on thermal and ionization properties}

As shown in Table~\ref{cons_table}, the thermal pressure remains consistent within uncertainties for X-ray components 3-5, despite their electron temperatures varying by a factor of 6. Furthermore, according to the {\tt tpho} model, the temporal variations of electron temperature due to changes of the source luminosity are subtle, accounting for less than 10\% of the mean value within each observation. Therefore, it is plausible to suggest that these components likely maintain stable pressure equilibrium within a shared environment (e.g., \citealt{chelouche2005, goosmann2016}). To achieve and sustain this equilibrium, it is necessary for the characteristic sound crossing time within the region to be shorter than its dynamical timescale, allowing for efficient propagation of pressure disturbances within the system.

On the other hand, the UV absorber components exhibit significantly lower gas pressure, which is approximately two or three orders of magnitude lower than the X-ray components. This contrast stems from a combination of differences in both the electron temperature and gas density. The pressure equilibrium observed in X-rays is not applicable to the UV component 1a, which exhibits a pressure approximately ten times higher than that of component 1b. Some of these UV components might represent cold gas with low ionization levels, undergoing transient heating and ionization from a cold neutral state \citep{gabel2005, tim2019}.

We present in Table~\ref{cons_table} the recombination time of Fe ions for each component, employing an approximate calculation based on charge state changes (see Eq. 5 in \citealt{l2023}). Our analysis reveals that components 3 and 4 exhibit recombination times of no more than a couple of days, while component 5 likely has a recombination time of a day or even shorter. These three components, especially component 5, show potential as candidates for investigating variability through timing analysis. According to Figure~\ref{fig:piontpho}, component 5 contributes significantly to the population of species ranging from \ion{Fe}{IX} to \ion{Fe}{XIII}, constituting a substantial portion of the Fe UTA. This finding aligns with the result of \citet{krongold2005}, who reported significant variability in the Fe UTA by utilizing a differential spectral technique on two HETGS spectra taken one month apart. Meanwhile, these timescales should still be in line with the report of \citet{behar2003}, where the Fe UTA was found to be relatively constant within few days, despite an increase in continuum flux. 

As presented in Table~\ref{cons_table}, the UV warm absorbers reported in \citep{gabel2005} exhibit longer recombination times, ranging from approximately $1.4$ to $3.5 \times 10^{6}$ seconds. This finding indicates that ion species with different levels of ionization may exhibit variations on distinct timescales. Considering that the peak source variation occurs within the range of $10^5$ to $5 \times 10^{6}$ seconds \citep{m2005}, which coincides with the observed range of recombination times, it is probable that many of the narrow absorption UV and X-ray components are in a non-equilibrium state, where they exhibit a delayed response to changes in the ionizing SED.

\subsection{Constraints on kinematics}
\label{4.4}

In order to provide a further understanding of the physical properties involved, we assess the observed densities in light of two useful constraints: the geometrically determined lower density limit, and the boundary between an escaping outflow and a failed wind. The lower density limit is derived from the assumption that the thickness of the outflowing cloud ($\Delta R$) is equal to the distance from the central supermassive black hole ($R$). In this scenario, the lower limit is given by $n_{\rm low} = L / (\xi \Delta R^2) = \xi N^2_{\rm H} / L$, where $\xi$ represents the ionization parameter, $N_{\rm H}$ denotes the column density, and $L$ corresponds to the average ionizing luminosity. The boundary between escaping and failed outflow is determined by the condition in which the outflow can barely overcome the gravitational potential when its radial velocity reaches $\sqrt{2 G M_{\rm BH}/r}$, where $G$ denotes the gravitational constant, $M_{\rm BH} = 2.8 \times 10^7$ $M_{\odot}$ represents the observed black hole mass \citep{bentz2021}, and $r$ corresponds to the distance. Accordingly, the density boundary can be expressed as $n_{\rm esc} = L v^{4}/(4 G^{2} M^2_{\rm BH} \xi)$, where $v$ represents the observed radial velocity of each component shown in Table~\ref{tab_pion}. 

The above estimation does not account for the transverse velocity component. It might be worth considering that some absorbers may have been moving perpendicular to our line of sight, as suggested by the decrease in the radial velocity of the UV absorbers reported in \citet{gabel2003}. Taking into account the possibility that the X-ray components also have a Keplerian velocity component of $\sqrt{G M_{\rm BH}/r}$, the boundary condition is modified to $n^{\ast}_{\rm esc} = L v^{4}/(G^{2} M^2_{\rm BH} \xi)$. The actual boundary would lie somewhere between $n_{\rm esc}$ and $n^{\ast}_{\rm esc}$.

Figure~\ref{fig:hd_v} show that the observed densities of X-ray absorber components 3-5 surpass $n_{\rm low}$ by a significant amount. Meanwhile, these densities all agree within uncertainties with $n_{\rm esc}$ and $n^{\ast}_{\rm esc}$, implying an intriguing possibility that the warm absorber components are being expelled from the system at velocities close to the escape velocity. It appears that these wind components possess just enough kinetic energy to exit the gravitational potential of the black hole and dissipate their energy in the interstellar medium. Assuming uniform cloud with spherical geometry, the present results show that the kinematic luminosities associated with these components range from approximately 10$^{30}$ to several 10$^{31}$ W. These values are significantly lower than the intrinsic AGN luminosity, which is estimated to be $\sim$ 6 $\times$ 10$^{36}$ W \citep{m2017}. As a result, it is unlikely that these warm absorbers play a significant role as an energy source of AGN feedback.

Based on the above estimates for the location and densities of X-ray components 3-5, it is plausible that these components occupy an intermediate distance range of $0.3-0.6$~pc, characterized by overlapping density and thermal pressure ranges. They might be influenced by a common stream of gas that moves at the escape velocity. Hence, it is possible that the combined spatial extents of these components, along with the media between them, could manifest as a geometrically thick region, where radiative transfer across different layers becomes potentially significant \citep{sadaula2023}. The present {\tt tpho} model does not account for time-dependent radiative transfer in such a thick layer. As a result, the current results may exhibit biases that could be addressed through an improved version of the spectral model.

\subsection{Atomic physics}
\label{4.5}

\citet{ballhausen2023} incorporated systematic uncertainties intrinsic in atomic data into their photoionization modeling for analyzing the {\it Chandra} HETGS data of NGC~3783. They showed that adjusting the radiative transition data for individual absorption lines brought a marginal $\sim$5\% improvement in the fit statistics. However, even with these modifications applied to the lines, the global fit quality remained formally unacceptable. They pointed out that the primary discrepancies between the data and the model likely stem from unaccounted astrophysical effects or uncertainties related to instrumental calibration.

Our analysis shown in \S\ref{sec:3.1} demonstrates that adopting our {\tt tpho} model yields a $\sim 6$\% improvement in the fit statistics with respect to the initial {\tt pion} fit. This implies that neglecting the time dependence in photoionization for this dataset caused a moderate impact on the fit statistics, which is, on the whole, comparable to that introduced by inaccuracies in atomic data. Despite this improvement, the final fit statistics with the present {\tt tpho} model is still not ideal. The major data-model discrepancies remain unresolved, which might originate from a combination of instrumental factors as well as unaddressed astrophysics, such as radiative transfer across layers and complex outflow velocity profiles.

\subsection{Future prospect}

\begin{figure}[!htbp]
\centering
\resizebox{1.0\hsize}{!}{\includegraphics[angle=0]{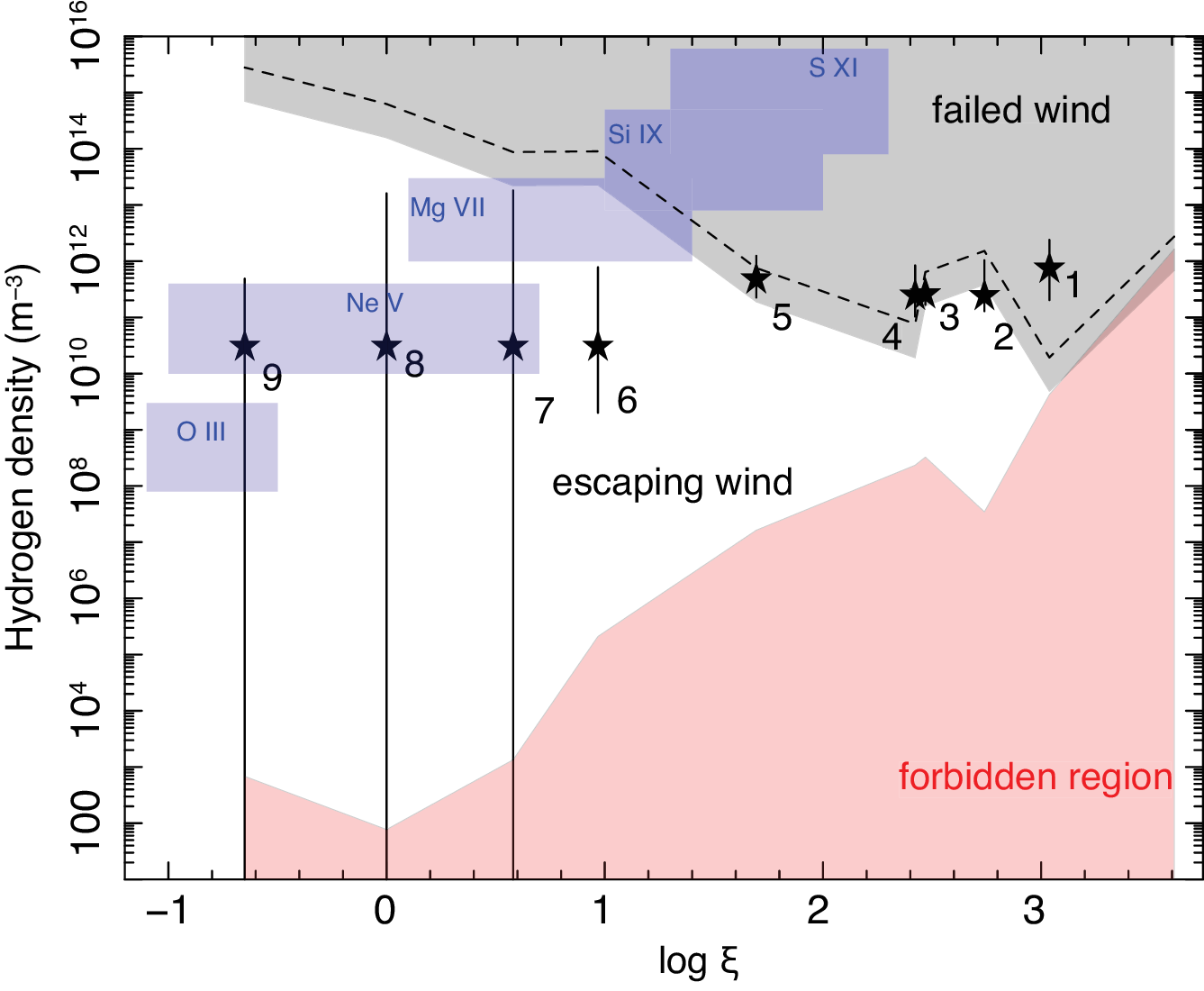}}
\caption{Same as Figure~\ref{fig:hd_v} but for the expected results from a simulation of 200~ks exposure with {\it Athena} X-IFU. }
\label{fig:athena}
\end{figure}

The HETGS spectra in our current study provide limited constraints on the densities of highly ionized absorbers. The planned X-ray Integral Field Unit (X-IFU, \citealt{barret2018}) on the Athena X-ray observatory \citep{nandra2013} will offer a substantial increase in effective area and spectral resolution with respect to the current instruments. This advancement will enable more precise and detailed spectroscopic investigations of warm absorbers, including the highly ionized components. To demonstrate this potential, we present a simulation based on the {\tt tpho} modeling framework.

The X-IFU spectrum is generated through a simulation based on the best-fit model obtained from observation 1, which includes nine {\tt tpho} components for the warm absorbers. The densities of components 1-5 are assigned the values presented in Table~\ref{cons_table}, while the densities of components 6-9 are assumed to be the same as that of component 1a observed in the UV \citep{gabel2005}. We incorporate a simulated lightcurve as shown in Figure~\ref{fig:lclong} into the calculation. The integrated time is set to 200~ks to capture the expected variability on a timescale of $\sim 10^5$~s. The simulation is built on the available response files of X-IFU, which incorporate a spectral resolution of 2.5~eV.

The density constraint is illustrated in Figure~\ref{fig:athena}. The most significant improvement will be observed in component 1 and 2, where the density can be determined with an accuracy of $\leq 50$\%. Components 3-5 will have density constraints achieved by the 200~ks X-IFU that are approximately twice better than the 900~ks HETGS. We will obtain a marginal constraint on component 6, and only upper limits for components 7-9. As shown in Figure~\ref{fig:athena}, the UV and X-ray metastable line diagnostics using, e.g., {\it Arcus} \citep{smith2022}, will serve as excellent complementary tools for these lowly ionized components.

The above result represents only a part the absorption study possibilities for AGNs using future instruments. The spectral analysis using the {\tt tpho} method has its limitations, as it necessitates a relatively long exposure to accumulate a high-quality time-integrated spectrum. On the other hand, timing analysis provides an alternative approach that is less constrained, since it does not require a comprehensive fitting of the full spectrum. The timing approach encompasses techniques employing Fourier timing \citep{silva2016} and coherence analysis \citep{j2022} to study short-timescale variability, as well as utilizing the density-delay relation for longer timescales \citep{l2023}. According to \citet{j2022}, the coherence modeling alone could potentially yield a density accuracy of $\leq 70-80$\% with X-IFU for outflows in a typical narrow-line Seyfert 1 AGN. Furthermore, \citet{l2023} suggests that the density-delay relation could be applied effectively when the source density falls within the range of approximately $10^{10}$ to $10^{13}$ m$^{-3}$ in NGC~3783. By leveraging the strengths of all available tools, we will be able to maximize our capability to determine the density and distance of the warm absorbers in AGNs.

\begin{acknowledgements}

SRON is supported financially by NWO, the Netherlands Organization for
Scientific Research. 
\end{acknowledgements}

\bibliographystyle{aa}
\bibliography{main}

\end{document}